 \documentclass[final,3p,times]{elsarticle}

\usepackage{epsfig}

\usepackage{amssymb}

\usepackage{sidecap}
\usepackage{caption}
\usepackage{subfigure}
\usepackage{url}

\usepackage{ulem}
\usepackage{xcolor}

\journal{Physica A}

\begin{document}

\begin{frontmatter}



\title{Finite-size effects in the rough phase of the 3d Ising model}


\author{Walter Selke\corref{cor1}}
\ead{selke@physik.rwth-aachen.de}
\cortext[cor1]{Correspondig author}

\address{Institut f\"ur Theoretische Physik, RWTH Aachen University, 
Germany}

\begin{abstract}
Using Monte Carlo simulations, finite-size effects of interfacial
properties in the rough phase of the Ising on a cubic lattice with
$L\times L\times R$ sites are studied. In particular, magnetization profiles
perpendicular to the flat interface of size L$\times$R are studied, with $L$
being considerably larger than $R$, in the (pre)critical temperature range. The
resulting $R$-dependences are compared with predictions of the standard
capillary-wave theory, in the Gaussian approximation, and with a field theory
based on effective string actions, for $L$=$\infty$.
\end{abstract}


\end{frontmatter}



\section{Introduction}
\label{S1:Introduction}

In the three-dimensional Ising model, the interface roughening transition
occurs at the temperature $T_r$, well below the bulk critical temperature,
$T_c$. The interface separates two regions with, predominantly, oppositely
oriented spins, $S_i=+1$ or $S_i= -1$, with $i$ denoting the lattice sites. In
the thermodynamic limit, at temperatures below $T_r$, the sharp interface
has a finite width, while above, $T_r$, i.e. in the rough phase, the interface
width diverges.

The existence of a roughening transition has been predicted about seventy
years ago \cite{Burton1949,Frank1951}. A rather simple theoretical description
is based on a Gaussian approximation of the energy needed for forming the area
of the interface \cite{Buff1965}. That description of the long-wavelength
fluctuations of the interface is usually called the capillary-wave theory
(CWT), see, e.g., the review by Gelfand and Fisher \cite{Gelfand1990} and the
fairly recent article by K\"opf and M\"unster \cite{KM2008}.

In this contribution, new Monte Carlo simulations related to the roughening
transition in 3d Ising models will be presented. Pioneering articles on that
transition have been published, especially, by Dietrich Stauffer and coauthors
\cite{Stauffer1983,Stauffer1990}. Cubic lattices with $L\times L\times R$ sites
have been studied, where the, at $T=0$, flat interface consists of $L\times L$
sites. $R$ has been fixed. Then, originally, the roughening transition has been
found to take place at $T_r/T_c \approx 0.56$ \cite{Stauffer1983}. Furthermore,
increasing $L$, the interface width, $W_i$, has been observed to diverge, in the
rough phase, logarithmically with $L$. This behavior is closely related to 
the Kosterlitz-Thouless transition of the 2d XY model
\cite{Stauffer1983,Stauffer1990,Kosterlitz1972,Mon1988}.

More recently, numerical estimates for $T_r$ and $T_c$ have been
improved. Nowadays, the, presumably, best estimates are $k_BT_r/J= 2.4537...$
\cite{Hasenbusch1996} and $k_BT_c/J= 4.51152...$ \cite{Landau2018}.
Obviously, the first Monte Carlo estimate for the roughening
transition \cite{Stauffer1983} has been confirmed rather well.

In this article, finite-size effects in the rough phase of the 3d Ising model
are studied. Again, we shall consider lattices with $L\times L\times R$ sites.
Now, however, the, at $T=0$, flat interface comprises $L\times R$ spins, with
$L$ being ''considerably larger'' than $R$, in contrast to the situation
outlined above. Accordingly, the aim will be to determine the R-dependence of
the interface width $W_i$. Numerical results will be compared to theoretical
predictions on the R-dependence in the limit L = $\infty$. On the one side,  
the finite-size behavior may be described in the framework of the Gaussian
approximation of the CWT, on the other side, there are more sophisticated
variants using concepts of string theory \cite{Caselle1994,Delfino2019}.

The outline of the article is as follows: In Section 2, the model and the
simulation method as well as related predictions of pertinent theories will be
sketched. Then, numerical results will be presented and compared to the
asymptotics suggested by the theories. Finally, conclusions will be given.


\section{Model, method, and theoretical predictions}
\label{S2:Modmet}

We study the nearest-neighbor Ising model with ferromagnetic couplings, $J$,
between neighboring spins, where $S_i= +/-1$ at site $i$. We consider cubic
lattices with $L \times L \times R$ spins in the $x$-, $y$- and $z$-directions.
The lattice constant is set equal to one. The interface is introduced by
fixing the boundary spins in the top and bottom $x$-$y$-layers, each
comprising $L \times L$ sites. In these two layers, at distance $R$, the spins
are fixed in the state -1 for $x<0$, while the spins are in the state +1 for
$x>0$. For $x=0$, the boundary spins are left free, setting them formally
equal to zero. At the other boundaries of the lattice, free boundary conditions
are used. Thence, in the ground state, at $T=0$, a flat interface, of area
$L \times R$, runs between the axes $x=0$ on the top and bottom layers, see
\cite{Delfino2019}.

Obviously, interesting properties of the interface follow from the
magnetization profile $m(x)$ in the central $x$-$y$-layer, $z=0$, of the
lattice, taking also $y$=0. At $T=0$, for the flat interface, $m(x)$ is
a step function, with $m(x)$= -1 for $x<0$, and $m(x)$= 1 for $x>0$. A
typical example for the magnetization profile in the rough phase is shown
in Fig. 1, as obtained in Monte Carlo simulations.
\begin{figure}
\resizebox{0.9\columnwidth}{!}{%
  \includegraphics{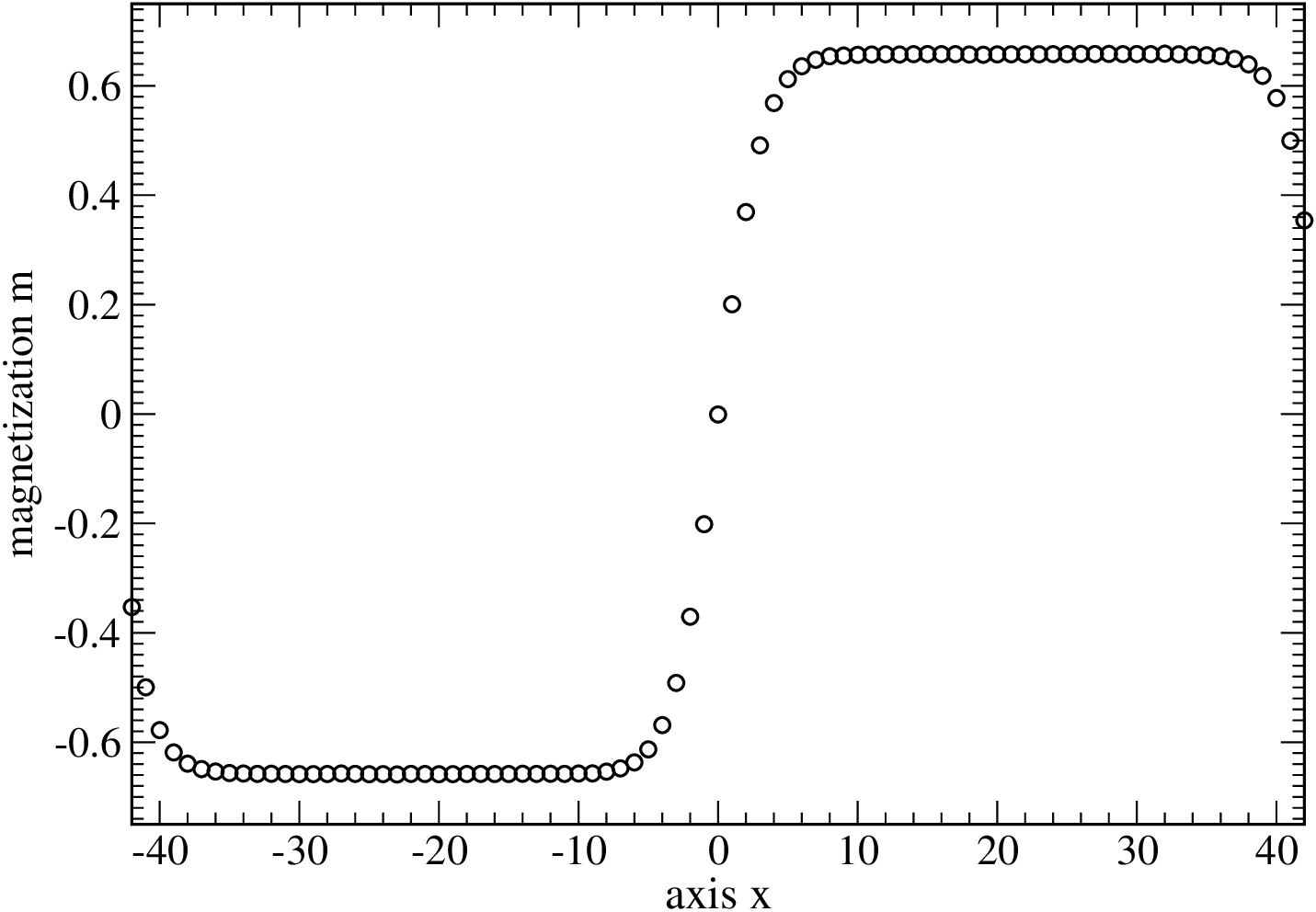}
}
\caption{Magnetization profile $m(x)$ at $k_BT/J= 4.2$ for a lattice with
  $L$=83 and $R$=21.}
\label{fig:1}
\end{figure}

Indeed, the magnetization profile, $m(x)$, plays a crucial role in the
theories describing finite-size behavior for the interface of this model, in
the case of $L= \infty$ and sufficiently large values of $R$.

In the CWT, the Hamiltonian is given by the change of the interface area due
to long-wavelength fluctuations. Then, the important term depends on
$\sqrt{1+|m'(x)|^2}$. In the Gaussian approximation, this expression is
replaced by the term in leading order of the Taylor expansion, being
proportional to $m'(x)^2$. For the geometry discussed above, the magnetization
profile, $m(x)$, is given by the error function
\begin{equation}
 m(x) \propto erf(\eta x)
\end{equation}
with $\eta \propto \sqrt{1/lnR)}$ at fixed temperature. Accordingly, the slope
of $m'(x)$ at the origin $x$=0 is given by
\begin{equation}
 m'(0) \propto \sqrt{(1/lnR)}
\end{equation}
The interface width, $W_i$, follows from $m(x)$ by the second moment
of its derivative $m'(x)$. Then, the finite-size behavior, as a function of the
length $R$, has the form
\begin{equation}
 W_i \propto lnR
\end{equation}
with the proportionality factor depending on temperature \cite{Caselle1996}.

A different $R$-dependence of the slope and the interface width has been
predicted by a recent string theory \cite{Delfino2019}.
This theory is supposed to be valid for $R$ being much larger than the
correlation length in the (pre)critical region of the model. In particular,
one obtains
\begin{equation}
 m'(0) \propto \sqrt{(1/R)}
\end{equation}
and 
\begin{equation}
  W_i \propto R
\end{equation}
To test the two conflicting theories, we do not concentrate on critical
phenomena very close to the bulk transition, so that standard Metropolis Monte
Carlo simulations seem to be suitable \cite{Binder2005}. To compare simulation
results with the theoretical predictions, $L$ is chosen to be ''considerably
larger'' than $R$. As stated above, in these theories, $L$ is assumed to
be infinite, and the theories describe dependences on the distance $R$.

Note, that previous Monte Carlo data, for magnetization profiles, $m(x)$, as
well as related energy profiles agreed nicely with predictions of the
string theory \cite{Delfino2019}. However, a systematic simulation study of
finite-size effects is still missing.

Specifically, we, mainly, performed Monte Carlo simulations at fixed
temperature, $k_BT/J$= 4.2, i.e. at about 0.93 $T_c$. We varied $R$ from 3 to
125, with $L$ increasing from 79 to 363. Typically, averages over at least
four, up to eight independent runs were taken. The typical length of each run
was $10^6$ to $10^7$ Monte Carlo steps per site. We computed magnetization
profiles, $m(x)$, along the axis $y= z= 0$. Then the resulting slope at $x=0$,
$m'(0)$, and the interface width, $W_i$, were plotted against the
$R$-dependences suggested by the two theories, see equations (2-5).


\section{Results}
\label{S3:Results}

To ensure that comparison of the simulation data to the theories assuming $L$
being infinite, is reasonable, we checked the dependence of the slope $m'(0)$
on the ratio $r$= $R/L$ for various fixed values of $R$. For $R$ ranging from
3 to 51, the slope was observed to increase monotonically when decreasing $r$.
When increasing $R$, the maximum, corresponding, presumably, to the case $r$=0,
was reached already at fairly large ratios. Indeed, for $R > 7$, the maximal
slope was approached closely already for $r$ being about 0.3-0.35
(then, $R$ is larger than the bulk correlation length of the Ising
model at the chosen temperature, $k_BT/J= 4.2$). In this way, we may quantify
the requirement, stated above, for comparing simulation results to the
theories, that $L$ is ''considerably larger'' than $R$. In fact, all
the data shown in Figs. 2 and 3 satisfy this condition. In fact, for $R < 11$,
the ratio $r$ was chosen to be even much smaller than 0.3.

\begin{figure}
\resizebox{0.75\columnwidth}{!}{%
  \includegraphics{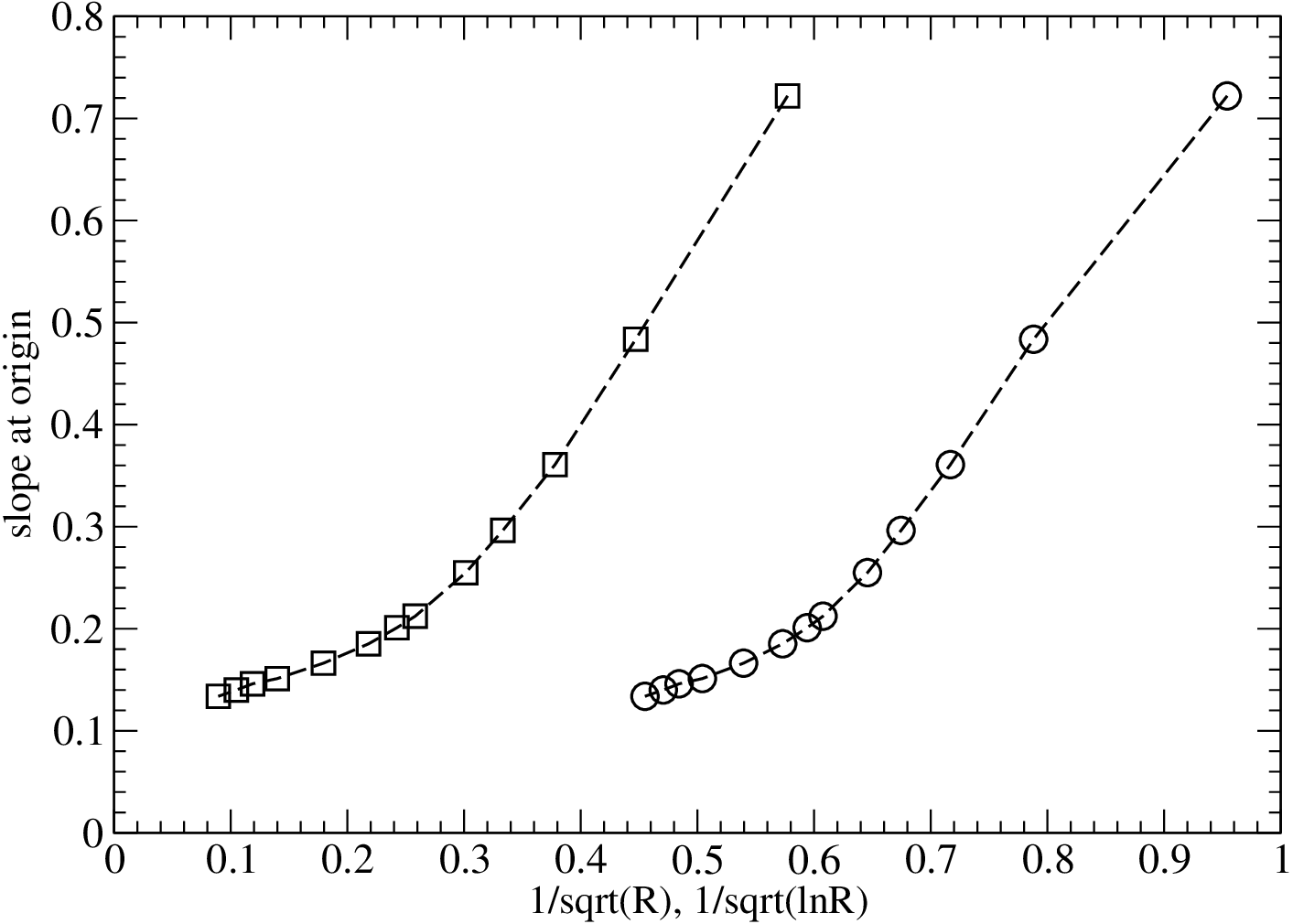}
}
  \caption{Slope of the magnetization profile at the origin, $m'(0)$, equation
    (6), plotted against $1/\sqrt{lnR}$ (circles) and against $1/\sqrt{R}$
    (squares). $R$ varies from 3 to 125, with the ratio $r=R/L$ being always
    smaller than 0.35 }
\label{fig:2}
\end{figure}

Results of the systematic Monte Carlo simulations for the interface width,
$W_i$, and for the slope of the magnetization profile at the origin, $m'(0)$,
are depicted in Figs. 2 and 3. The focus is, at $k_BT/J= 4.2$, on the
$R$-dependence, choosing sufficiently large values of $L$, to allow for
comparison with the capillary wave theory, in the Gaussian approximation, and
with the string theory.

\begin{figure} 
\resizebox{0.75\columnwidth}{!}{%
  \includegraphics{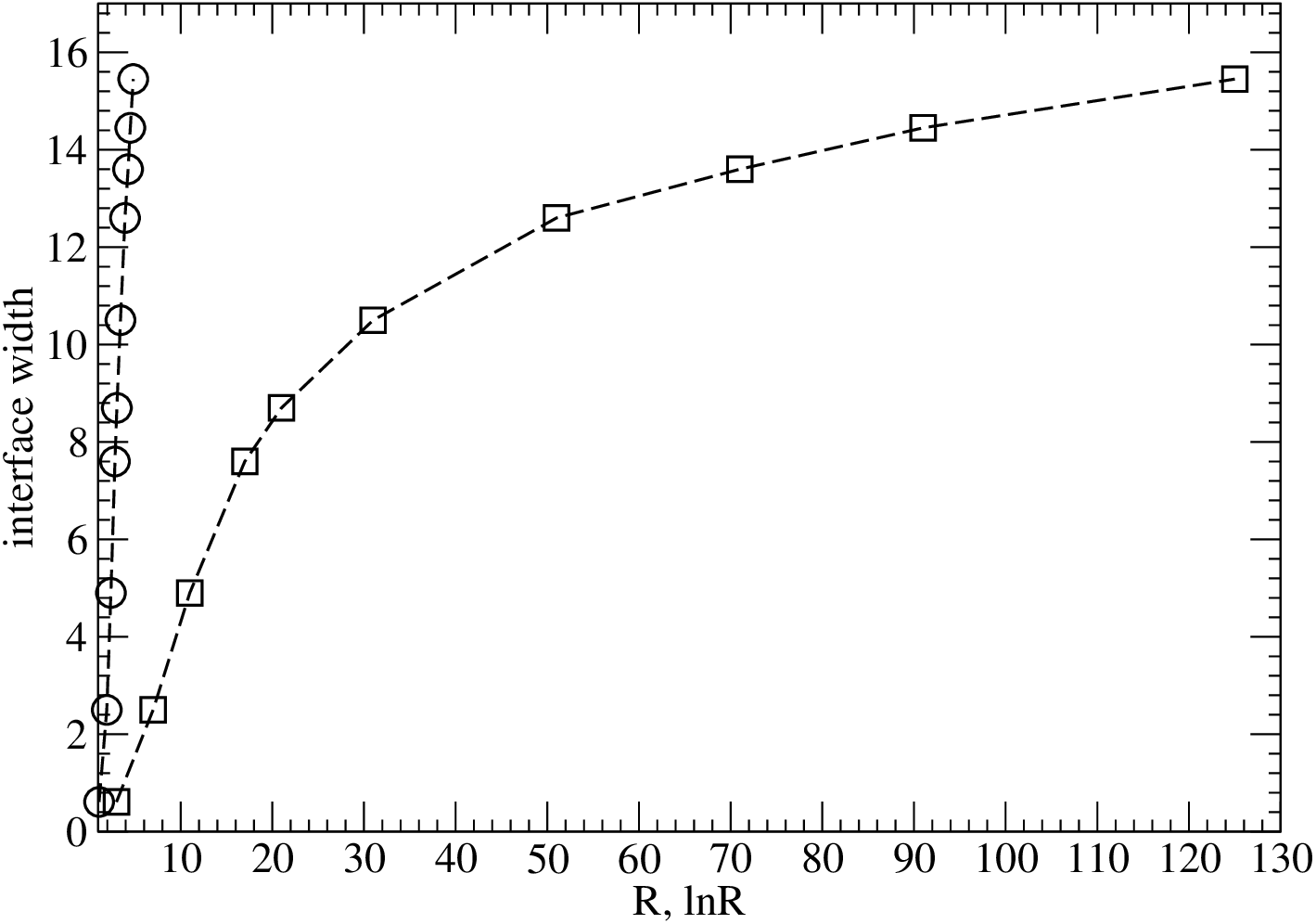}
}
\caption{Interface width $W_i$, equation (7), plotted against $lnR$
  (circles) and $R$ (squares), for the same lattices as in figure 2 }
\label{fig:3}
\end{figure}

The two theories predict different asymptotics for the $R$-dependences of
interfaces in the limit $L=\infty$, see equations (2-5). Accordingly, we
analyzed both the slope of the magnetization profile at the origin, eqs. (2)
and (4), as well as the interface width, eqs. (3) and (5). Due to the discrete
lattice structure, the slope may be approximated by

\begin{equation}
  m'(0)= (m(1) - m(-1))/2
\end{equation}

The interface width may be defined by

 \begin{equation}
  W_i =  F \sum\limits_{x} m'(x+1/2) \times (x+1/2)^2/\sum\limits_{x}m'(x+1/2)
\end{equation}
with $m'(x+1/2)=m(x+1)-m(x)$. The sums run between the two plateaus of the
magnetization profile, where $m'$ approaches zero, see figure 1. Of course, the
regions close to the (free) boundaries are ignored, to avoid their influence.

The simulation results for the slope and for the interface width are depicted in
Figs. 2 and 3. In both cases, the $R$-dependence has been plotted against the
asymptotic forms suggested by the capillary wave and the string theories, as
discussed above.
  
For the slope $m'(0)$, figure 2, we observe, for sufficiently large distances
$R$, linear dependences both in 1/$\sqrt{lnR}$, as predicted by CWT, as well
as in 1/$\sqrt{R}$, as predicted by the string theory. However, simple
extrapolation to $R=\infty$ would give only in case of the scale proposed by
the CWT a vanishing slope in the limit $R=\infty$, while in case of the scale
according to the string theory, a limiting slope of about 0.1 seems to result,
when extrapolating the simulation data. This finding will be addressed again
later in this section.

The interface width $W_i$ is depicted in figure 3, plotted against
$lnR$, in accordance with the CWT, and against $R$, in accordance with the
string theory. Here, a linear dependence seems to set in already at
moderate values of $R$, when testing the CWT. On the other hand, the
simulation data for $W_i$ appear to approach, if at all, the asymptotic
form, suggested by the string theory, only for larger values of $R$. Again, the
discussion on this topic will be postponed to the Conclusions.- Note
that similar behaviors have been found when estimating the interface width by
the $R$-dependence of the distance from the origin to the position, at which
the magnetization profile reaches 0.9 $m_p$, with $m_p$ being the plateau
value of magnetization $m(x)$, see figure 1. Of course, this quantity provides
a rather crude measure of the interface width.

Note that a less elaborate MC simulation study for lattices of moderate sizes,
ranging from $R$= 7 up to 51, in the rough phase at a lower temperature,
$k_BT/J =3.4$, shows good agreement with the CWT, both for the slope of the
magnetization profile at the origin and for the interface width, see,
equations (2) and (3). Interestingly, in the string case, the magnetization
slope tends to a rather large non-zero value, of about 0.35, when
extrapolating the MC data linearly to the limit $R= \infty$, compare to
figure 2.

On the other hand, we performed  preliminary simulations at a temperature
closer to $T_c$, $k_BT/J =4.4$, with $R$ ranging from 15 to 45 lattice units.
At this temperature, a straightforward linear extrapolaton of the MC data for
the magnetization slope, plotted versus 1/$\sqrt{R}$, leads to a very small
value, of less than 0.01, in the limit $R= \infty$. In fact, this behavior may
signal possible agreement with the string theory. Of course, more detailed
computations, possibly, even closer to $T_c$ and for sufficiently large
lattices, are desirable.


\section{Conclusions}
\label{S4:Conclusion}
 
Systematic and rather extensive Monte Carlo simulations, especially, at
$k_BT/J =4.2$, have been performed to study finite-size effects of interface
properties in the rough phase of the three-dimensional Ising model. The
interface properties are related to the magnetization profile $m(x)$.

Lattices of $L \times L \times R$ sites are considered, with flat interfaces
being of size $L \times R$. Simulation results are compared to predictions of
two theories, the standard capillary wave theory, CWT, and a string theory.
These theories, in the limit of $L= \infty$, lead to different asymptotics of
finite-size dependences on $R$. According to the string theory, in the
(pre)critical regime, rather unusual finite-size effects, for sufficiently
large distances $R$, may occur.

In particular, we studied the slope of the magnetization profile in the center
of the interface, $m'(0)$, as well as the interface width, $W_i$. For both
quantities, the finite-size dependence, of $R$, agrees quite well with the
predictions of the CWT for a wide range of values of $R$. At first
sight, more pronounced deviations of the simulation data from the asymptotics
following from the string theory are observed. Nevertheless, one has to be
cautious in ruling out that theory. Most importantly, the string theory is
expected to hold, asymptotically, for large values of $R$, in the
(pre)critical regime, i.e. below, but sufficiently close to $T_c$
\cite{Delfino2019}. Thence, further simulations at higher temperatures seem to
be desirable. Presumably, numerical algorithms reducing critical slowing down,
in particular, cluster-flip updates \cite{Binder2005}, will be useful.

In addition, both theories are based on continuum descriptions. In contrast,
in the Monte Carlo simulations, discrete lattices are studied.

In any event, open questions remain. In particular, the connection between the
two theories needs to be clarified. Furthermore, the role of temperature, in
the rough phase below the critical point, by going closer to $T_c$, should be
elucidated.


\section*{Acknowledgements}
\label{S5:Acknowledgements}

Useful comments by G. M{\"u}nster are gratefully acknowleged.- I
gladly remember numerous inspiring and helpful discussions with Dietrich
Stauffer, for about four decades. DS worked successfully on a wide range of
topics, starting in traditional Statistical Physics, using various numerical
and analytic methods. He liked to communicate, in a clear and humorous way,
his profound insights to experts as well as to beginners.





\section*{References}

\end{document}